\documentclass[twocolumn,twoside,fleqn,showpacs,showkeys,pra,aps,10pt,superscriptaddress,final]{article}

\usepackage{hep}
\graphicspath{{figs/}}      %定义图的存放路径

\def\refname{References}

\def\journalname{??}

\def\@pacs@name{PACS numbers: }%
\def\@keys@name{Keywords: }%

\def\Dated@name{Dated: }%
\def\Received@name{Received }%
\def\Revised@name{Revised }%
\def\Accepted@name{Accepted }%
\def\Published@name{Published }%
\def\address{\replace@command\address\affiliation}%
\def\altaddress{\replace@command\altaddress\altaffiliation}%

\definecolor{orangec}{cmyk}{.24,.91,.96,.18}
\definecolor{orangecc}{cmyk}{.24,.94,.96,.18}
\definecolor{oorangec}{cmyk}{.8,.2,.5,.4}
\definecolor{ooorangec}{cmyk}{1,.9,0.08,.04}      %四色

\definecolor{orangec}{cmyk}{.15,.7,.96,.0}
\definecolor{orangecc}{cmyk}{.15,.7,.96,.0}
\newcommand{\ODDBOX}{\noindent\raisebox{-13.3cm}{\includegraphics{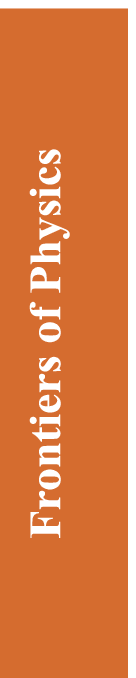}}}
\newcommand{\EVENBOX}{\noindent\raisebox{-13.3cm}{\includegraphics{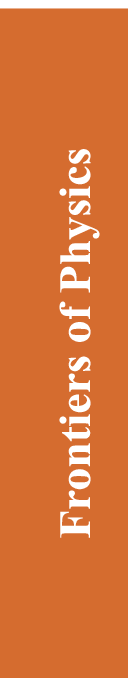}}}

\newfont{\yihao}{cmb10 at 18pt}
\newcommand{\Yihao}{\fontsize{18pt}{13.5pt}\selectfont}
\newcommand{\xiaosi}{\fontsize{11.5pt}{13.5pt}\selectfont}         % 小四字体   %\陈清华增加

\newfont{\xbt}{cmb10 at 12pt}

\def\frontmatter@title@format{%
    \centering%
    \usefont{T1}{fradmcn}{m}{n}\yihao}%
\def\@keys@name{{\color{ooorangec}\bf Keywords~~}}%
\def\@pacs@name{{\color{ooorangec}\bf PACS numbers~~}\vspace{2mm}}%
\def\frontmatter@authorformat{\vspace{5mm}\centering\bf}%

\newcommand{\catchline}[2]{
	{\vspace*{-16.4mm}\small%
	\noindent #1\\%
	\noindent #2\\[-2mm]%
    {\color{orangec}{\rule{\textwidth}{.5pt}}}\\[2mm]
    {\color{orangec}{\Yihao\textbf{\textsc{\Papertype}}}}\\[6mm]
    }\relax\par
}

\newcommand{\doi}[1]{DOI {#1}}
\renewcommand{\title}[1]
{\vspace*{-5mm}\begin{center}
{\Yihao\bf #1}
\end{center}
}

\renewcommand{\author}[1]
{\vspace*{0mm}
\begin{center}
{\bf #1}
\end{center}
}
\newcommand{\add}[1]{\begin{center}{\small\it #1}\end{center}}

\newcommand{\abs}[1]{
\begin{center}
\parbox[t]{156mm}{\noindent\color{oorangec}#1}
\end{center}}

\newcommand{\keywords}[1]{
\begin{center}
\parbox[t]{156mm}{\noindent{\bf\color{ooorangec}Keywords}\ \ #1}
\end{center}}

\newcommand{\pacsnumbers}[1]{
\begin{center}
\parbox[t]{156mm}{\noindent{\bf\color{ooorangec}PACS numbers}\ \ #1\vspace*{5mm}}
\end{center}}

\newcommand{\acknowledgements}[1]{\vspace*{4mm}\noindent{\renewcommand{\baselinestretch}{1.05}\footnotesize{\color{ooorangec}\bf Acknowledgements}\quad{#1}}}

\def\journalname{??}
\def\volumenumber#1{\gdef\@volumenumber{#1}}%
\def\@volumenumber{}%
\def\issuenumber#1{\gdef\@issuenumber{#1}}%
\def\@issuenumber{}%
\def\volumeyear#1{\gdef\@volumeyear{#1}}%
\def\@volumeyear{}%

\renewcommand\thesection{\arabic{section}}
\renewcommand\thesubsection{\arabic{section}.\arabic{subsection}}
\renewcommand\thesubsubsection{\arabic{section}.\arabic{subsection}.\arabic{subsubsection}}

\titleformat{\section}[hang]{\color{ooorangec}\vspace*{-1.2mm}\titlerule\vspace{1mm}\large\usefont{T1}{fradmcn}{m}{n}\xbt}{\thesection}{1em}{}
\titlespacing{\section}{0mm}{8mm}{5mm}

\titleformat{\subsection}{\normalfont\normalsize\color{ooorangec}}{\thesubsection}{1em}{}
\titlespacing{\subsection}{0mm}{5mm}{3mm}

\titleformat{\subsubsection}{\normalfont\normalsize\it\color{ooorangec}}{\thesubsubsection}{1em}{}
\titlespacing{\subsubsection}{0mm}{3mm}{3mm}

\titlecontents{section}
[0em] {\normalfont} {\thecontentslabel\quad} {\thecontentslabel}
{\titlerule*[.7pc]{}\contentspage}

\titlecontents{subsection}
[1.5em] {\normalfont} {\thecontentslabel\quad} {\thecontentslabel}
{\titlerule*[.7pc]{}\contentspage}

\titlecontents{subsubsection}
[4em] {\normalfont} {\thecontentslabel\quad} {\thecontentslabel}
{\titlerule*[.7pc]{}\contentspage}

\usepackage[small]{caption2}%去掉图注后昌号%%%[hang,footnotesize]表示图注在图号后自控对齐
\newlength{\halfpagewidth}
\divide\halfpagewidth by 1

\usepackage{hyperref}
\usepackage{mathrsfs}
\usepackage{ulem}

\begin{document}

\newcommand{\Papertype}{\sc Research article} % 文章类型
\def\volumeyear{2020} % 出版��?
\def\volumenumber{???}  %{15(5)}  %{15(5)} % ��?
\def\issuenumber{??}%{54501} % 编号
\def\journalname{Front. Phys.} %期刊��?
\newcommand{\doiurl}{???} %{10.1007/s11467-020-0966-4} % DOI链接
\newcommand{\allauthors}{Fen LYU, Ya-Ping Li, Shu-Jin Hou et al.}
\twocolumn[
\begin{@twocolumnfalse}
\catchline{\journalname~\volumenumber,~\issuenumber~(\volumeyear)}{\doi{\doiurl}} %Please ignore
\thispagestyle{firstpage}

%\begin{strip}

\title{Self-organized Criticality in Multi-pulse Gamma-Ray Bursts}

%\author{Xiangru Li$^{1*}$, Woliang Yu$^{2}$, Xilong Fan$^{3**}$,G. Jogesh Babu$^{4}$}
\author{Fen LYU$^{1,2}$, Ya-Ping Li$^{3*}$, Shu-Jin Hou$^{4}$, Jun-Jie Wei$^{1}$, Jin-Jun Geng$^{6,7**}$, Xue-Feng Wu$^{1,5***}$}

\add{1. Purple Mountain Observatory, Chinese Academy of Sciences, Nanjing 210023, China\\
	2. University of Chinese Academy of Sciences, Beijing 100049, China\\
	3. Theoretical Division, Los Alamos National Laboratory, Los Alamos, NM 87545, USA\\
	4. College of Physics and Electronic Engineering, Nanyang Normal University, Nanyang, Henan 473061, China\\
	5. School of Astronomy and Space Science, University of Science and Technology of China, Hefei, Anhui 230026, China\\
	6. School of Astronomy and Space Science, Nanjing University, Nanjing 210023, China\\
	7. Institute of Astronomy and Astrophysics, University of T\"ubingen, Auf der Morgenstelle 10, D-72076,T\"ubingen\\
%\add{1. School of Computer Science, South China Normal University, Guangzhou 510631, China\\
%2. School of Mathematical Sciences, South China Normal University, Guangzhou 510631, China\\
%3. School of Physics and Technology, Wuhan University, Wuhan, Hubei 430072, China\\
%4. Pennsylvania State University, University Park PA, 16802, USA\\
%Corresponding author.\ E-mail: $^*$lixiangru@scnu.edu.cn, $^{**}$xilong.fan@whu.edu.cn\\
Corresponding author \ E-mail: $^{*}$leeyp2009@gmail.com, $^{**}$gengjinjun@nju.edu.cn,$^***$ xfwu@pmo.ac.cn\\}
%Received March 25, 2020; Accepted May 06, 2020}

\abs{The variability in multi-pulse gamma-ray bursts (GRBs) may help to reveal the mechanism of underlying processes from the central engine. To investigate whether the self-organized criticality (SOC) phenomena exist in the prompt phase of GRBs, we statistically study the properties of GRBs with more than 3 pulses in each burst by fitting the distributions of several observed physical variables with a Markov Chain Monte Carlo approach, including the isotropic energy $E_{\rm iso}$, the duration time $T$ and the peak count rate $P$ of each pulse. Our sample consists of 454 pulses in 93 GRBs observed by the CGRO/BATSE satellite. The best-fitting values and uncertainties for these power-law indices of the differential frequency distributions are: $\alpha^d_{E}=1.54 \pm 0.09$, $\alpha^d_{T}=1.82_{-0.15}^{+0.14}$ and $\alpha^d_{P}=2.09_{-0.19}^{+0.18}$, while the power-law indices in the cumulative frequency distributions are: $\alpha^c_{E}=1.44_{-0.10}^{+0.08}$, $\alpha^c_{T}=1.75_{-0.13}^{+0.11}$
	and $\alpha^c_{P}=1.99_{-0.19}^{+0.16}$. We find that these distributions are roughly consistent with the physical framework of a Fractal-Diffusive, Self-Organized Criticality (FD-SOC) system with the spatial dimension $S=3$ and the classical diffusion $\beta$=1. Our results support that the jet
	responsible for the GRBs should be magnetically dominated and magnetic instabilities (e.g., kink model, or tearing-model instability)
	lead the GRB emission region into the SOC state.}


\pacsnumbers{98.70.Rz, 02.70.Rr}

\vspace*{-6mm}

\end{@twocolumnfalse}
]

\section{%1~
Introduction} \label{sec:intro}

Gamma-ray bursts (GRBs) are extremely energetic events occurring at the cosmological distance. The observed GRB lightcurves usually consist of several pulses characterized by highly temporal variabilities. It is well accepted that the prompt gamma-ray emission is generated by internal dissipation processes while the later afterglow is produced through the shock wave interacting with the surrounding medium. There is a consensus that long GRBs originate from the collapse of massive stars \citep{1992ApJ...397..570M}, while short GRBs are from mergers of two compact objects such as binary neutron stars or black hole-neutron star binaries \citep[e.g.][]{
	1986ApJ...308L..43P,2006Sci...311.1127D,
	2011MNRAS.415.2495M,
	2017ApJ...848L..14G}. The nature of GRB's central engine has remained mysterious. The central engine is popularly supposed to be a black hole surrounded by a hyper-accreting disk \citep{1999ApJ...524..262M} or a millisecond magnetar \citep{2006Sci...311.1127D}, but remains uncertain from case to case.
On the other hand, the presence of X-ray flares after the prompt gamma-ray emission indicates the GRB central engine may not cease all activities after the main burst phase
\citep[e.g.][]{2005Sci...309.1833B,
	2006Sci...311.1127D,
	2006ApJ...646..351L,
	2006ApJ...642..354Z,2019ApJ...884...59L}.

The self-organized criticality (SOC) phenomena are ubiquitous, and they are commonly observed in many astrophysical processes, such as solar flares \cite{2020FrPhy..1534601G}, magnetospheric substorms, lunar craters, pulsar glitches, and fast radio bursts \cite{2017JCAP...03..023W,2019arXiv190103484L,
	2019arXiv190311895Z}, etc (see \citep{2014ApJ...782...54A,
	2016SSRv..198...47A} for a review). The general definition of SOC is a critical state of a nonlinear energy dissipation system that is slowly and continuously driven towards a critical value of a system-wide instability threshold, producing scale-free, fractal-diffusive, and intermittent avalanches \citep{2014ApJ...782...54A,1987PhRvL..59..381B,1988PhRvA..38..364B}. The SOC phenomena could be identified and diagnosed by analyzing the power-law or power-law like frequency distributions of relevant scale-free parameters \cite{2010ApJ...717..683A,2012A&A...539A...2A,2014ApJ...782...54A,2016SSRv..198...47A}.

Some SOC phenomena in GRBs have been discussed in the literature. The frequency distributions of energy/waiting time\footnote{ The general definition of waiting time is the time interval between two subsequent bursts.} for GRB X-ray flares exhibit
power-law tails very similar to those of solar X-ray flares \cite{2013NatPh...9..465W} indicating that the central engine activity of GRBs might be a magnetic reconnection-driven SOC system like that happens on the Sun.
It was later revealed that statistical properties of energy, duration, peak flux, and waiting time for X-ray flares from different sources (the Sun, GRBs, Swift J1644+57, Sgr A$^{\ast}$, and M87) could be wholly explained by the SOC model \cite{2015ApJS..216....8W}. Also, the statistical similarities between those sources imply that all of the X-ray flares are consistent with magnetic reconnection events. The dimensionality of the SOC process in Sgr A$^{\ast}$ X-ray flares is discussed \cite{2015ApJ...810...19L}. The waiting time distributions of both $\gamma$-ray pulses and X-ray flares of GRBs were systematically analyzed and argued that
it is not proper to use the SOC mechanism as the interpretation to GRBs \cite{2015ApJ...801...57G}.
However, Yi et al. \cite{2016ApJS..224...20Y} comprehensively studied GRB X-ray flares observed by the Swift satellite, and their results supported the SOC phenomena in GRBs.
Moreover, it is found that these statistical properties are similar to those of the X-ray flares, which indicates that GRB optical flares and X-ray flares may share a common physical origin \cite{2017ApJ...844...79Y}. Besides, the power density spectra (PDSs) of GRB prompt lightcurves (with a power-law index value of $\sim -5/3$) indicate the bursts themselves have a self-similar temporal structure \citep{2000ApJ...535..158B}.

Some studies suggest that X-ray flares and the gamma-ray prompt emission may share a common origin, i.e., relativistic jets \citep{2015ApJ...808...33U,2018ApJ...862..115G}. It is straightforward to ask whether the SOC phenomena exist in the GRB prompt emission and if so, whether the SOC behavior of multiple-pulses in the prompt emission is consistent with that of X-ray flares, and what causes this SOC behavior.
This motivates us further to investigate the statistical properties of multiple pulses to explore the SOC behavior in the main prompt emission of GRBs.

The most interesting statistical characteristics of the SOC phenomena are scale-free power-law or the  power-law-like size distributions (or frequency distributions) of the physical parameters of the system (\cite{2015ApJ...814...19A} and the references therein).

In this study, we investigate the statistical characteristics of GRBs with multiple pulses in their prompt $\gamma$-ray lightcurves. The structure of this article is as follows. We present the selection criterion for our GRB sample in Section 2. In Section 3, we study the frequency distributions of several physical quantities, including the energy, the duration time, and the peak count rates of the pulses of GRBs. Finally, in Section 4, we discuss the implication of the statistical results and summarize our conclusions. Throughout this work, we assume a flat $\Lambda$CDM universe with the cosmological parameters $\Omega_{m}=0.32$, $\Omega_{\Lambda}=0.68$, and $H_{0}=72\,{\rm km}\,{\rm s}^{-1}{\rm Mpc}^{-1}$.

\section{%2~
Sample and data reduction}\label{sec:Sample}
GRB lightcurves are highly complex and diverse, varying from a smooth single pulse to spiky multi-pulses.
The pulses in the prompt emission should be the imprints of activities of the GRB central engine.
Here, we investigate the statistical distributions of three physical variables of GRB pulses,
including the isotropic energy $E_{\rm iso}$, the duration time $T$, and the peak count rate $P$.

As the Burst and Transient Source Explorer (BATSE; \cite{1989BAAS...21..860F,1992Natur.355..143M}) onboard in the Compton Gamma-Ray Observatory (CGRO; \cite{1994ApJS...92..351G}) has a wider energy range and a lower photon detection threshold, we collect the GRB sample from BATSE, rather than other instruments like HETE-2, Swift, or Fermi.
From the BATSE GRB catalog \cite{2011ApJ...740..104H}, we select the bursts whose lightcurves have at least three pulses. For these bursts, their pseudo redshifts are obtained from the catalog \cite{2004ApJ...609..935Y},
which are estimated based on the $E_{p}$-luminosity relation.
Since spectral parameters are essential to derive the $k$-correction to the $E_{\rm iso}$ and $P$,
we choose the bursts that have spectral parameters in the BATSE 5B GRB spectral catalog \citep{2013ApJS..208...21G}.
With this selection criterion, we finally collect 454 pulses in a sample of 93 GRBs.

$k$-correction to burst energy in its cosmological rest frame should be considered in our calculations.
For a burst at redshift of $z$, the corrected isotropic energy is written as
\begin{equation}
E_{\mathrm{iso}}=\frac{4 \pi D_{L}^{2} F_{\gamma}}{(1+z)}\times\frac{\int_{30~\mathrm{keV} / (1+z)}^{10^4~\mathrm{keV} / (1+z)} E \times N(E) d E}{\int_{E_{\min}}^{E_{\max}} E \times N(E) d E} \operatorname{erg}
\end{equation}
where $F_{\gamma}$ is the pulse fluence, and $D_{L}$ is the luminosity distance, and the spectral function $N(E)$ is the empirical Band function \citep{1993ApJ...413..281B}. The spectral parameters for $N(E)$ could be referred from the catalog \cite{2011ApJ...740..104H}. Note that the pseudo redshifts of our sample were estimated based on the empirical luminosity relation, in which the integration of $k$-corrections factor is performed from 30 keV and $10^{4}$ keV \cite{2004ApJ...609..935Y}. To keep the consistence, we here adopt the same range of integration \cite{2002A&A...390...81A,2004ApJ...609..935Y}.
	 $E_{\min}$ and $E_{\max}$ are 25 keV and 300 keV, respectively, because the pulse fluence is observed in this energy range.

For the pulse duration $T$ in the source rest frame, it is $T = \omega/(1+z)$, where $\omega$ is the observed pulse width obtained from Table 1 in the catalog \cite{2011ApJ...740..104H}. Note that here $\omega$ is based on time intervals between times when the pulse intensity is $Ae^{-3}$, rather than $Ae^{-1}$, and $A$ is the pulse amplitude \citep{2005ApJ...627..324N}.
The pulse peak count rate $P$ (in unit of $\mathrm{counts}~\mathrm{s}^{-1}$) in the rest frame is calculated as
\begin{equation}
P=p_{64}
\times\frac{\int_{30~\mathrm{keV} / (1+z)}^{10^4~\mathrm{keV} / (1+z)}  N(E) d E}{\int_{E_{\min}}^{E_{\max}} N(E) d E}
\operatorname{cts} \mathrm{s}^{-1}
\label{eq:P}
\end{equation}
where $p_{64}$ is the peak flux in 64 ms bin.

\section{%3~
SOC analyses}\label{sec:analyses}

While an ideal power-law distribution function is commonly used in the standard SOC models,
	most observed frequency distributions of empirical data are not consistent with an ideal power law.
	Hence, we will use the thresholded power-law distribution function \cite{2015ApJ...814...19A} to analyze the data in our work.

In general, for the number of events N, the observed differential distribution could be well described with a so-called thresholded power-law distribution
\begin{equation}
N_{\mathrm{diff}} = \frac{{\rm d}N}{{\rm d}x} (x) \propto (x+x_0)^{-\alpha}, \ \ \ x_1\leq x \leq x_2,
\label{eq_diff}
\end{equation}
where $x_0$ is a constant by considering the threshold effects (e.g., incomplete sampling below $x_0$, background contamination), $x_1$ and $x_2$ are the minimum and maximum values of $x$, respectively. The uncertainty of the differential distribution is given by
\begin{equation}
\sigma_{\text {\rm diff}, i}=\sqrt{N_{\rm bin,i}} / \Delta x_{i}
\end{equation}
where $N_{\rm bin,i}$ is the event number in the $i$-th bin, and $\Delta x_{i}$ is the bin size. And for $\alpha \ne 1$, the corresponding cumulative distribution function (CDF) is written as
\begin{equation}
N_{\mathrm{cum}} (>x) = N_{\rm env}\times \left(\frac{(x_{2}+x_{0})^{1-\alpha}-(x+x_{0})^{1-\alpha}}{(x_{2}+x_{0})^{1-\alpha}-(x_{1}+x_{0})^{1-\alpha}}\right),
\label{eq_cum}
\end{equation}
where $N_{\rm env}$ is the total number of events. The uncertainty of the cumulative distribution in a given bin $i$ is estimated with
\begin{equation}
\sigma_{\mathrm{cum}, i}=\sqrt{N_{i}},
\end{equation}
where $N_{i}$ is the number of events of the bin.

For a fractal-diffusive system, the statistical properties of the event duration ($T$), event intensity ($P$) and event energy($E$) are critical for determining the SOC structures \citep{2014ApJ...782...54A}. In the following, we discuss the physical implication of the prompt pulse variables' differential distributions in the framework of FD-SOC model. For the fractal dimension of avalanches, the FD-SOC theory predicts the indices of differential distributions for $E_{\rm iso}$, $T$ and $P$ \citep{2014ApJ...782...54A} as
\begin{equation}
\begin{array}{l}\label{equ:square}
\alpha_{E}=1+(S-1)/(D_{S}+2/\beta)\\
\alpha_{T}=1+(S-1) \beta/2\\
\alpha_{P}=2-1/S\;,
\end{array}
\end{equation}
where $S$ is the Euclidean space dimensionality, $\beta$ is the diffusion parameter, $D_{S}$ is the mean value of a fractal dimension, which is calculated by $D_{S}\approx \frac{(1+S)}{2}$ \citep{2012A&A...539A...2A}.
%
%\textbf{Generally, there are two methods to explore SOC behavior usually. One is to explore the power-law index of some physical quantities (such as energy, peak flux, duration, waiting time) in the differential distribution (or the differential distribution), another is to explore the correlation relationships of these quantities to derive the statistical SOC parameters. But the latter usually causes a large certain about the theoretical value due to various factors (e.g., the detection ability limit, the small sample size).}

We explore the differential size distributions and the cumulative size distributions of observed physical variables (i.e., $E_{\rm iso}$, $T$, and $P$), and fit them with the theoretical model.	
We apply a uniformly logarithmic binning for the differential and the cumulative distributions. The method for determining the numbers of bins is similar to that in Li et al. \cite{2015ApJ...810...19L}. For the differential distributions, the bins that have no events are omitted.

We obtain the best fitting parameter $\alpha$ by minimizing the reduced $\chi_{\rm d.o.f}$ for the differential distribution function,
\begin{equation}
\chi_{\mathrm{d} . \mathrm{o.f}}=\sqrt{\frac{1}{\left(N_{x}-N_{\mathrm{par}}\right)} \sum_{i=1}^{N_{\mathrm{x}}} \frac{\left[N_{\mathrm{diff}}\left(x_{i}\right)-N_{\mathrm{diff}, \mathrm{obs}}\left(x_{i}\right)\right]^{2}}{\sigma_{\mathrm{diff}, i}^{2}}}
\end{equation}
%Normally, the uncertainties of the indices in the differential size distributions are a litte larger than those in the cumulative size distributions, the major reason is that the cumulative size distributions would contain favorably more events in each bin than the differential distributions. That's why we also explore the quantities in the cumulative size distribution.

 The $\chi_{\mathrm{d} . \mathrm{o.f}}$ for the cumulative distribution function $N(>x)$
\begin{equation}
\chi_{\mathrm{d} . \mathrm{o.f}}=\sqrt{\frac{1}{\left(N_{x}-N_{\mathrm{par}}\right)} \sum_{i=1}^{N_{x}} \frac{\left[N_{\mathrm{cum}}\left(x_{i}\right)-N_{\mathrm{cum}, \mathrm{obs}}\left(x_{i}\right)\right]^{2}}{\sigma_{\mathrm{cum}, i}^{2}}},
\end{equation}
where $N_{x}$ is the number of logarithmic bins, and $N_{\mathrm{par}}$ is the number of the free parameters.
$N_{\text {diff,obs}}\left(x_{i}\right)$ and $N_{\text {cum,obs}}\left(x_{i}\right)$
are the corresponding observed values for the differential distribution and the cumulative distribution, respectively.

\begin{figure*}
	\begin{center}
		\includegraphics[angle=0,scale=0.50]{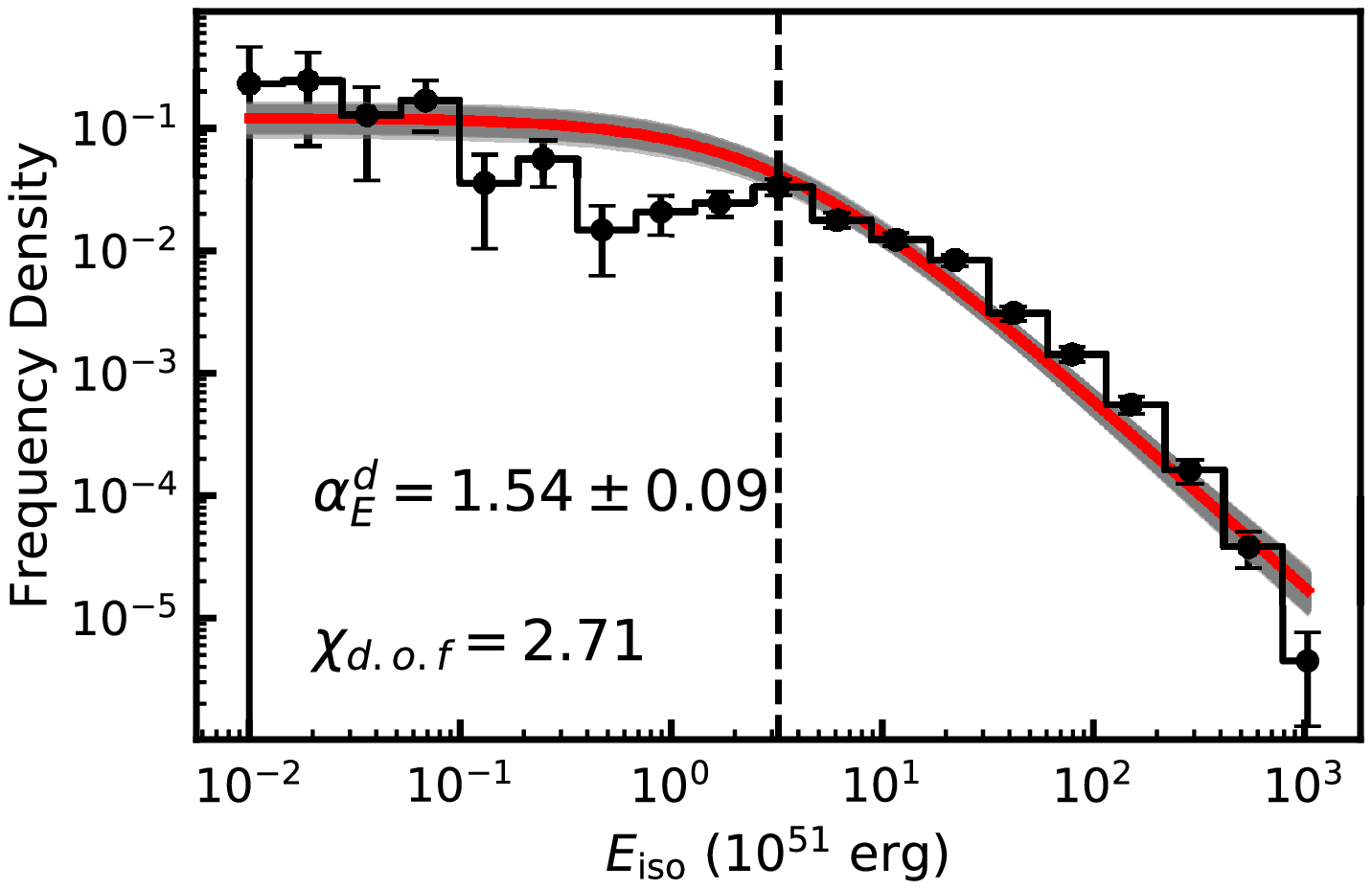}
		\includegraphics[angle=0,scale=0.50]{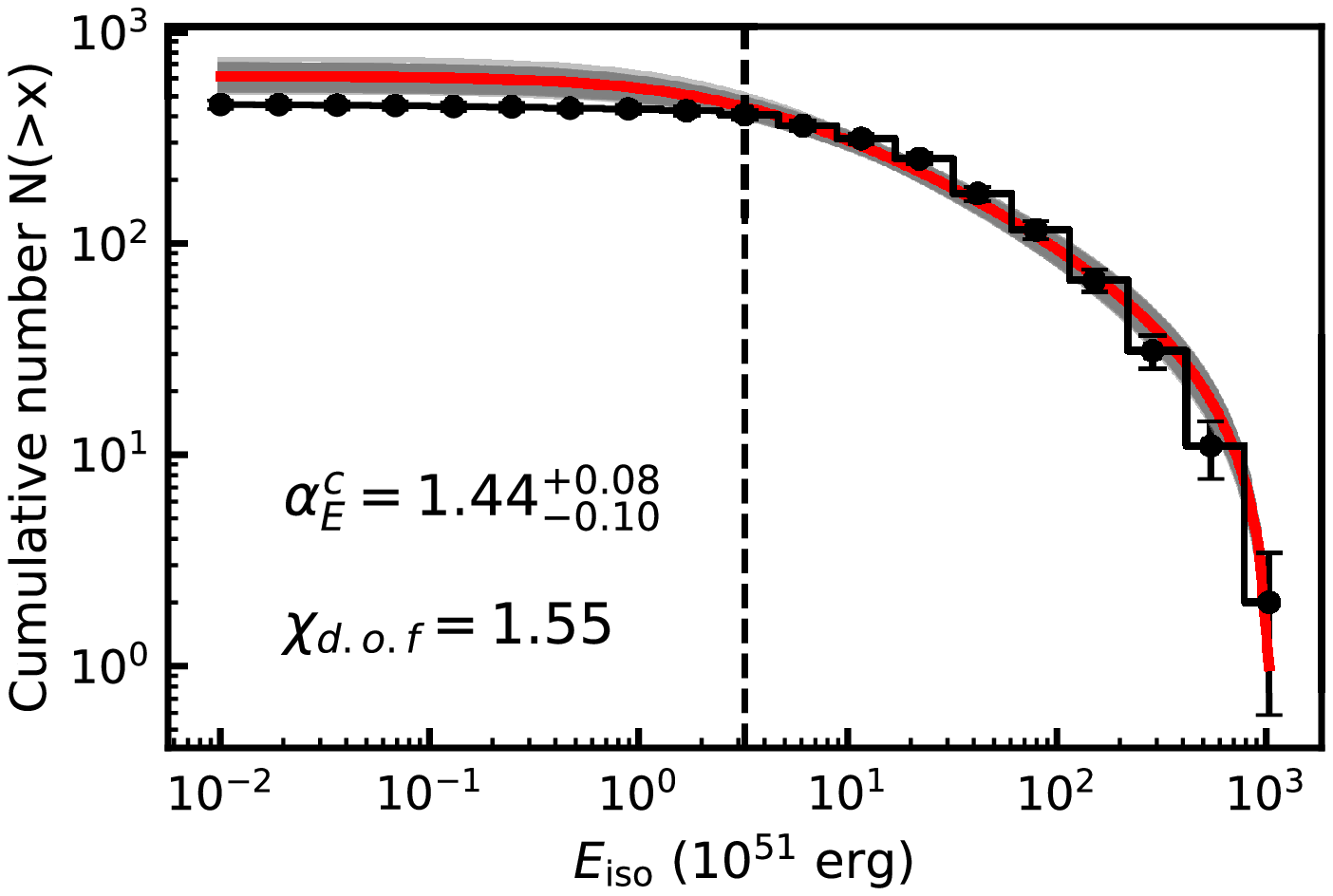}
		\caption{Left: the differential distribution of the pulse isotropic energy $E_{\mathrm{iso}}$; Right: the cumulative distribution of the pulse $E_{\mathrm{iso}}$.
			The vertical dashed line corresponds to the threshold of the isotropic energy, which is $x_{0}=3.22\times10^{51}$ erg. The red lines represent the best-fitting line using equations \ref{eq_diff} and \ref{eq_cum}. The maximum value of the pulse isotropic energy is  $1.03\times10^{54}$ erg.
			The gray shadow shows the $3 \sigma$ fitting range for the best-fitting line.}
		\label{Fig_1}
	\end{center}
\end{figure*}

\begin{figure*}
	\begin{center}
		\includegraphics[angle=0,scale=0.50]{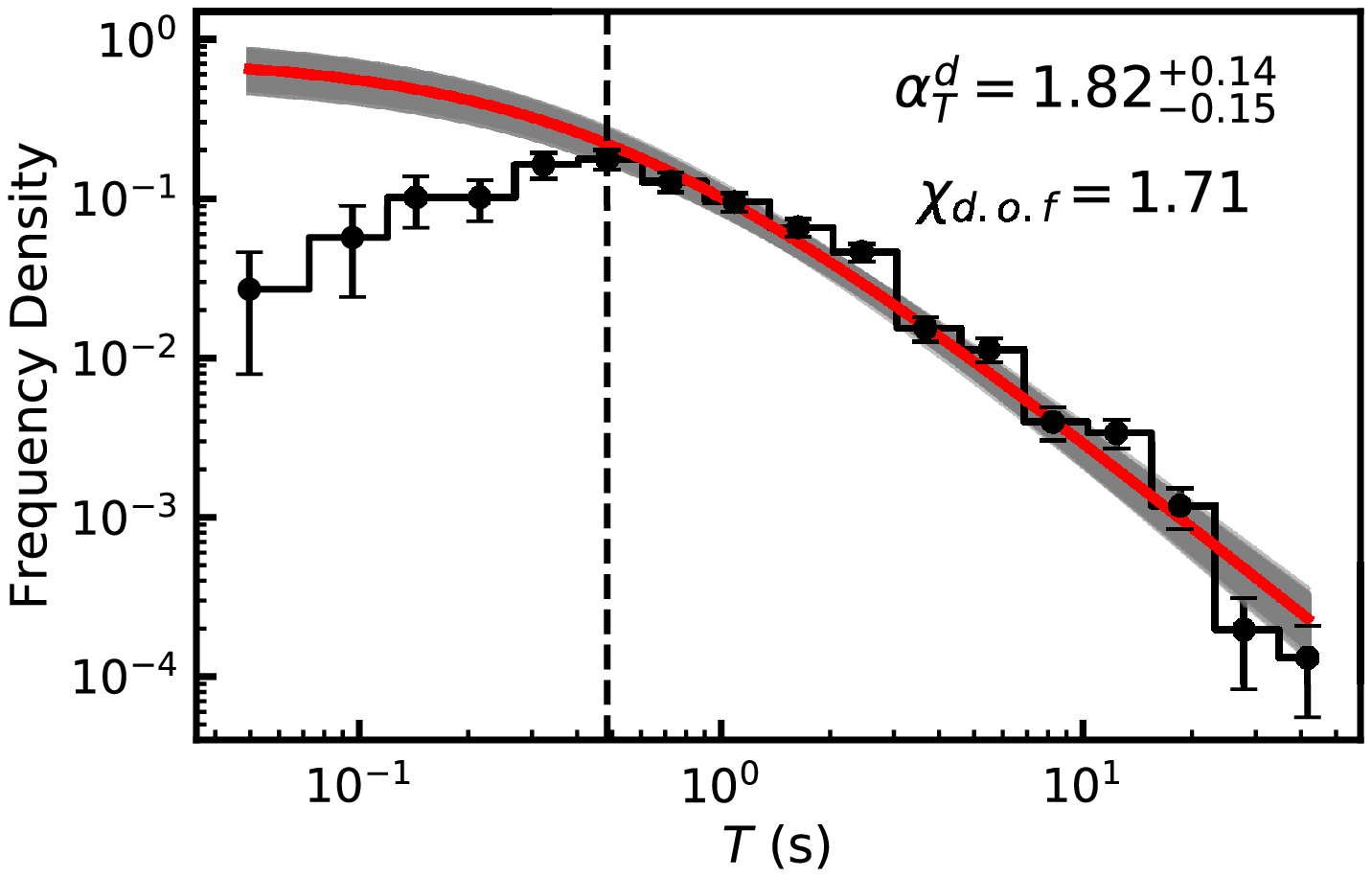}
		\includegraphics[angle=0,scale=0.50]{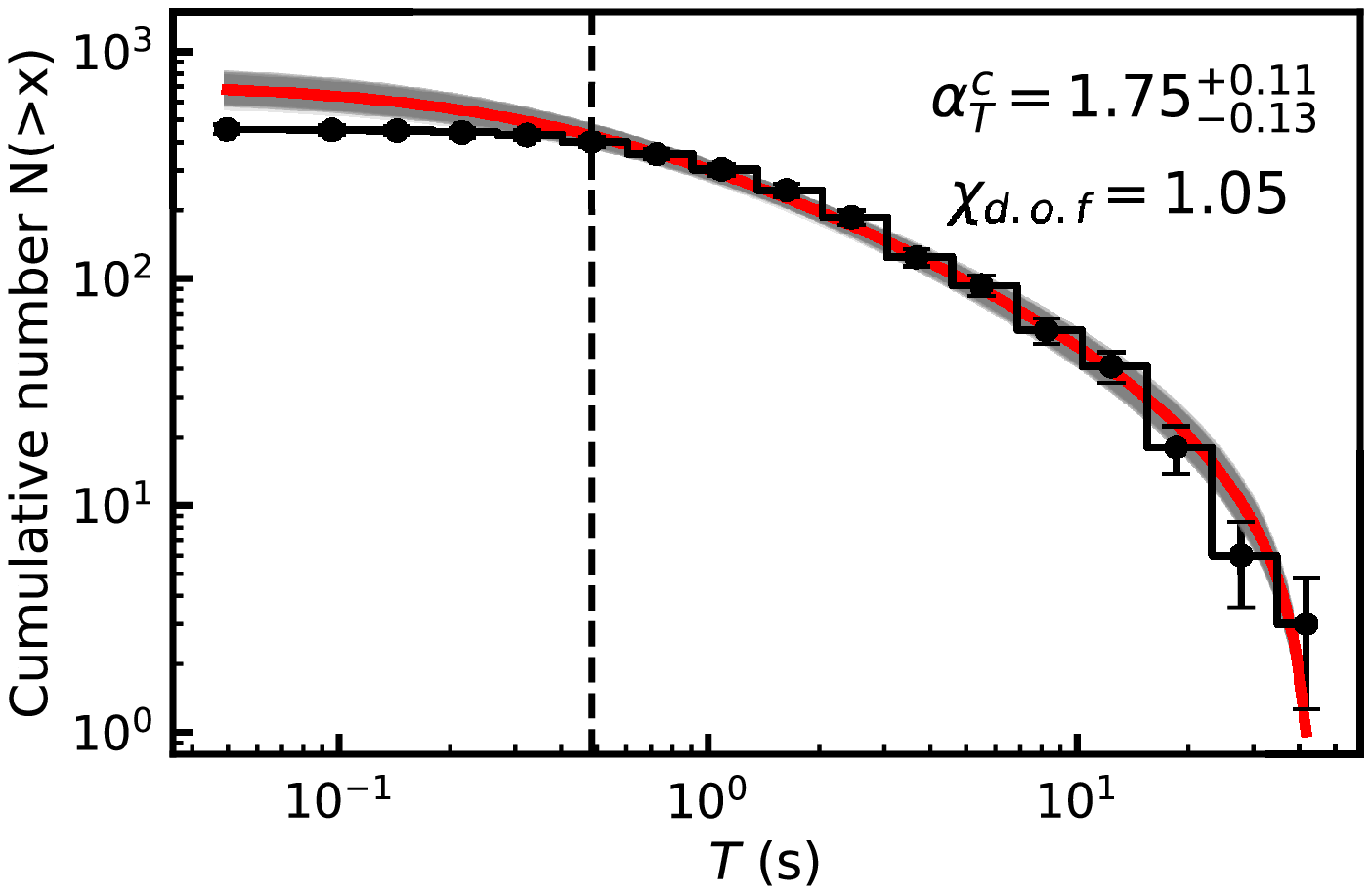}
		\caption{Left: the differential distribution of the pulse duration time $T$; Right: the cumulative distribution of the pulse $T$.
			The red line is the best-fitting line using equations \ref{eq_diff} and \ref{eq_cum},
			and the vertical dashed line corresponds to the threshold of the pulse duration, which is 0.48 s. The maximum value of the pulse duration is 41.72 s.
			The gray shadow shows the $3 \sigma$ fitting range for the best-fitting line.}
		\label{Fig_2}
	\end{center}
\end{figure*}

\begin{figure*}
	\begin{center}
		\includegraphics[angle=0,scale=0.50]{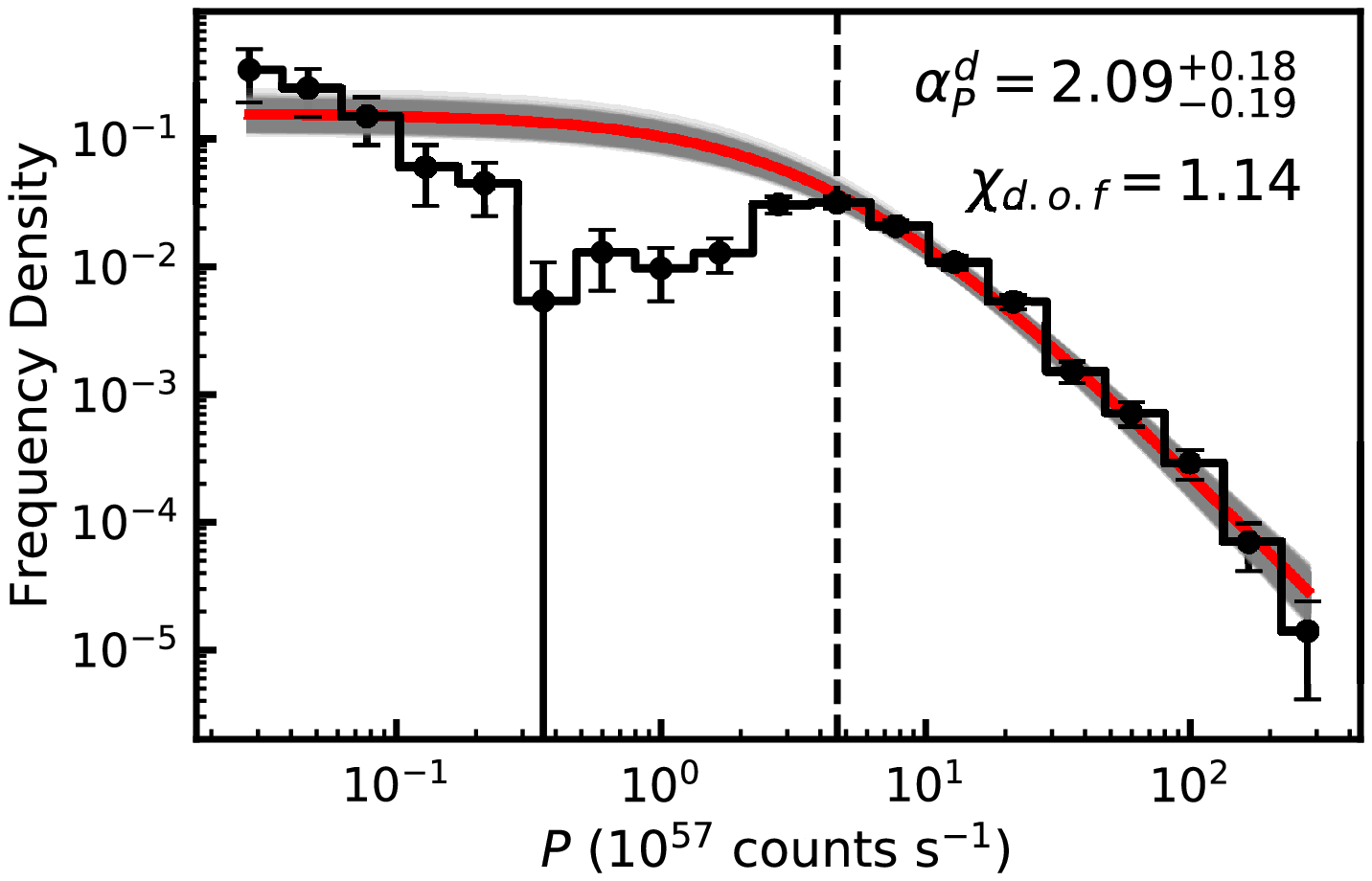}
		\includegraphics[angle=0,scale=0.50]{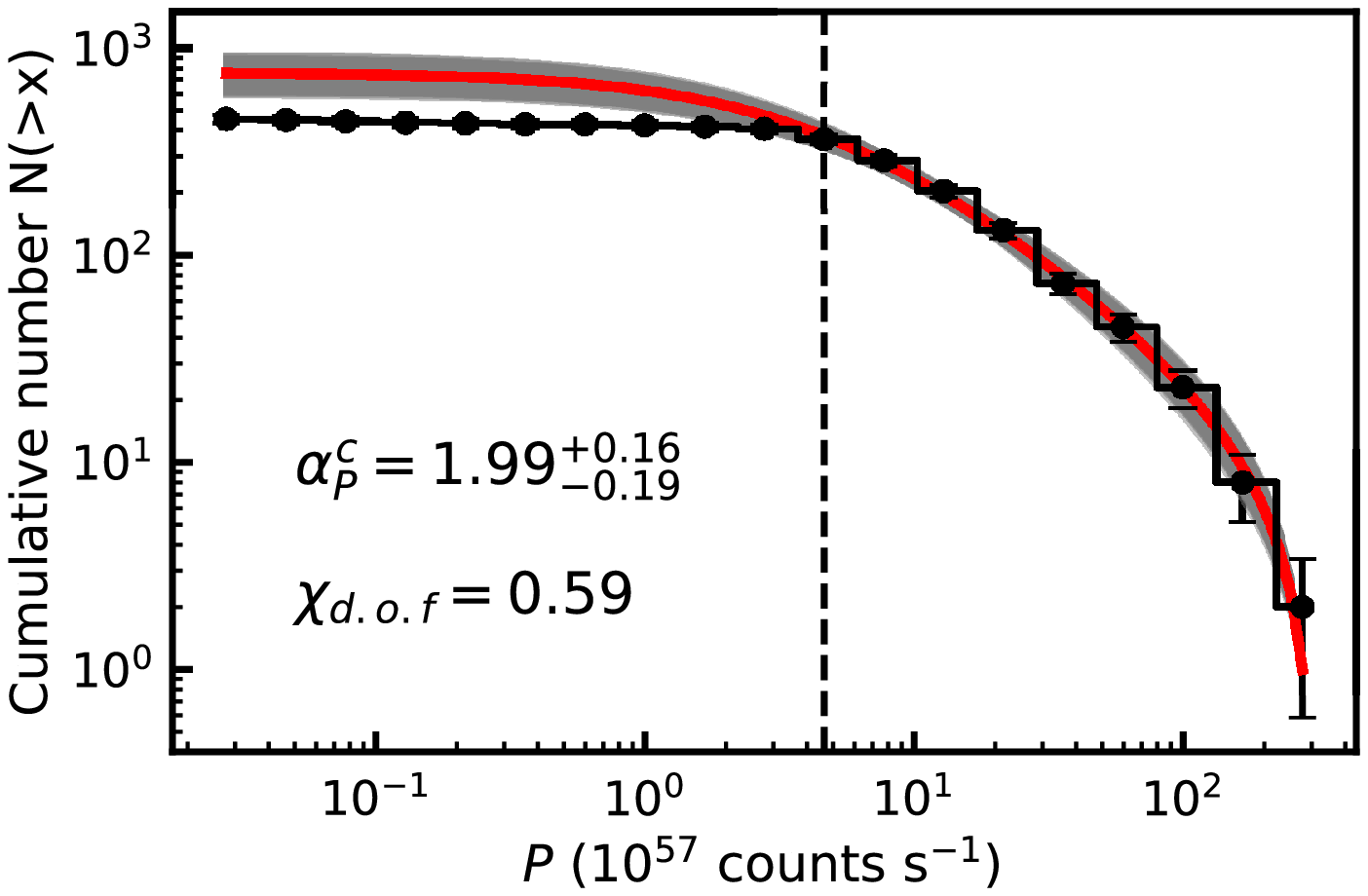}
		\caption{Left: the differential distribution of the pulse peak count rate $P$; Right: the cumulative distribution of the pulse $P$.
			The red line is the best-fitting line using equations \ref{eq_diff} and \ref{eq_cum},
			and the vertical dashed line corresponds to the threshold of peak count rate,
			which is about $4.63 \times 10^{57}$ counts $\rm s^{-1}$. The maximum value of the pulse peak count rate is $2.77 \times 10^{59}$ counts $\rm s^{-1}$.
			The gray shadow shows the $3 \sigma$ fitting range for the best-fitting line.}
		\label{Fig_3}
	\end{center}
\end{figure*}

In equations \ref{eq_diff}-\ref{eq_cum},  there are two crucial parameters, i.e., $\alpha$ and $x_0$. It has been revealed \cite{2015ApJ...814...19A} that the cumulative distributions turn out to be highly degenerate over $x_0$. Therefore, when taking $x_0$ as a free parameter together with $\alpha$ to fit the cumulative distributions, $\alpha$ could not be well constrained. Here we take the same $x_{0}$ as in the differential distribution with the same binning.
The position of emerging breakpoint shown in the differential distribution is taken as the value of $x_0$,
as shown in the left panels of Figs. \ref{Fig_1} - \ref{Fig_3}.
This is primarily based on the definition of $x_0$,
which was introduced to account for the deviation from an ideal power-law like behavior at the lower part of the differential distribution. The fitting parameters of these two kinds of differential distributions can be determined via the emcee algorithm ( Python implementation of Markov chain Monte Carlo fitting technique, \cite{2013PASP..125..306F}) if $x_0$ is fixed.

The differential distribution of the pulse $E_{\rm iso}$ is displayed in the left panel of Figure \ref{Fig_1} with $x_0 = 3.22 \times 10^{51}$ erg, and the power-law index fitted using Equation~(\ref{eq_diff}) yields $\alpha^{d}_{E} = 1.54 \pm 0.09$ ($\chi_{\mathrm{d} . \mathrm{o.f}}$=2.71\footnote{Note that the MCMC fitting and the calculation of $\chi_{\mathrm{d} . \mathrm{o.f}}$ are performed by considering the data above $x_{0}$ only}). The right panel of Figure \ref{Fig_1} illustrates the corresponding cumulative distribution. With the threshold $x_{0}$ determined by the differential distribution, the best-fitting power-law index of the cumulative distribution is $\alpha^{c}_{E}=1.44_{-0.10}^{+0.08}$ ($\chi_{\mathrm{d} . \mathrm{o.f}}$=1.55).

The differential distribution of the pulse duration time ($T$) for the $\gamma$-ray multi-pulses sample is displayed in the left panel of Figure \ref{Fig_2},
which gives the best-fitting index as $\alpha^{d}_{T}=1.82_{-0.15}^{+0.14}$ ($\chi_{\mathrm{d} . \mathrm{o.f}}$=1.71) given the threshold value to be 0.48 s.
The corresponding cumulative distribution is shown in the right panel of Figure \ref{Fig_2}, gives the best-fitting index as $\alpha^{c}_{T}=1.75_{-0.13}^{+0.11}$ ($\chi_{\mathrm{d} . \mathrm{o.f}}$=1.05).

Also, in Figure \ref{Fig_3}, it is shown that the differential distribution and the cumulative distribution of $P$ could be fitted with indices of
$\alpha^{d}_{P}=2.09_{-0.19}^{+0.18}$ ($\chi_{\mathrm{d} . \mathrm{o.f}}$=1.14), and $\alpha^{c}_{P}=1.99_{-0.19}^{+0.16}$ ($\chi_{\mathrm{d} . \mathrm{o.f}}$=0.59) respectively, given the threshold value to be $4.63\times 10^{57}$ counts $\rm s^{-1}$.

All the results above have quoted a 3$\sigma$ uncertainty for the best fitted power-law index.
The derived indices from the fits to both the differential and cumulative distributions are generally consistent within the error bars.
It is obvious from Figures 1$\sim$ 3 that the values of $\chi_{\mathrm{d} . \mathrm{o.f}}$ for the differential
size distributions are larger than those in the cumulative size distributions. The major reason is that
the cumulative size distributions would contain favorably more events in each bin than the differential distributions.

The choice of $x_{0}$ from the data itself would lead to the relatively large $\chi_{\mathrm{d} . \mathrm{o.f}}$=2.71 in the left panel of Figure 1 since $x_{0}$ is deeply affected by the effect of incomplete sampling and background contamination.

 In the framework of the FD-SOC model, we obtain
 $\alpha_{E}=1.5$, $\alpha_{T}=2.0$ and $\alpha_{P}$$\simeq$ 1.7 according to Equation ($\ref{equ:square}$) by taking the three-dimensional space ($S = 3$) and the classical diffusion ($\beta = 1$). Comparing to our results, it is found that the derived $\alpha^{d}_{E}$ and $\alpha^d_{T}$  are well consistent with the prediction of the FD-SOC model and $\alpha^{d}_{P}$ also roughly consistent with the model prediction. $P$ is derived from 64 ms peak flux $p_{64}$ (see equation \ref{eq:P}).
 However, $p_{64}$ is replaced by an averaged flux over the pulse duration for several long-timescale pulses in the catalog.
 Considering that the averaged flux over the pulse duration is relatively fainter than the flux determined in short time interval \citep{2015SCPMA..58.5575L},
 $\alpha^{d}_{P}$ is expected to be steeper than its predicted value.

%{\bf ***I think this paragraph should be removed*** Note that the uncertainties of the indices in the differential distributions are larger than those in the cumulative distributions. The major reason is that the cumulative distributions would contain favorably more events in each bin than the differential distributions. Also, the fitting lines deviate significantly from the data at the low end of all variables' cumulative distributions. The two main effects causing this deviation are: (1) an internal physical threshold of the underlying instability; (2) the incomplete sampling (limited instrument observation capability) of the smallest events below the threshold $x_0$. If this deviation is attributed to the intrinsic threshold, the threshold value could serve as an upper limit to the physical instability. On the other hand, the deviation from ideal power law at the high end of cumulative distribution is due to truncation effects at the largest events due to the finite system size.}

\section{%3
Discussions and conclusion}\label{sec:discuss}
In this paper, we have compiled 454 prompt $\gamma$-ray pulses in 93 GRBs observed by the CGRO-BATSE satellite within ten years. The reason why we select the mutiple-pulses dected by the CGRO-BATSE satellite as the only sample source because of its lower photon detection threshold (observed pulses are more distinct) compared with other detectors.
At the same time, the total number of GRBs in our sample is limited by the number that have pseudo redshifts which were served as a distance indicator.

By analyzing the temporal properties (i.e., $E_{\rm iso}$, $T$ and $P$) of these pulses, we presented that the differential and the cumulative thresholded differential distributions of GRB temporal properties could be well understood within the physical framework of a fractal-diffusive, self-organized criticality (FD-SOC) model, which is generally consistent with the previous works on the SOC behavior in GRB X-ray flares \citep{2013NatPh...9..465W,2015ApJS..216....8W,2016ApJS..224...20Y,2017ApJ...844...79Y}. However, the statistical results favor a spatial dimension of $S = 3$ of the SOC system for prompt properties, rather than a spatial dimension of $S = 1$
for the case of X-ray flares \citep{2013NatPh...9..465W}.

The positive thresholds of these physical parameters in three distributions (see Figures 1$\sim$ 3) suggest that the statistical results of SOC behavior among multiple pulses in the prompt emission are slightly contaminated by the event-unrelated background \citep{2015ApJ...814...19A}.

Pulses are building blocks of one GRB event. Due to the highly overlapping effect, it is difficult to identify the pulses from complex lightcurves. In this analysis, we adopt a sample from \citep{2011ApJ...740..104H} and present SOC analysis. Note that scale-free is a very important characteristic of a SOC system. Although extracting pulses from lightcurves highly depends on the empirical pulse model and the instrument threshold, our analysis based on the uniform criteria could give insights to evaluating the SOC nature of GRBs.

%\textbf{The two main effects causing this deviation are: (1) an internal physical threshold of the underlying instability; (2) the incomplete sampling (limited instrument observation capability) of the smallest events below the threshold $x_0$. If this deviation is attributed to the intrinsic threshold, the threshold value could serve as an upper limit to the physical instability.
%On the other hand, the deviation from ideal power law at the high end of cumulative size distribution is due to truncation effects at the largest events due to the finite system size. The SOC distribution is a heavy tail distribution indeed, and statistical properties in the differential size distributions are generally consistent with that in the corresponding cumulative distributions, so the incomplete sampling has little influence on it.}
Another effect is that the physical threshold of the instability of pulses in the prompt emission of GRBs. In a SOC system, owing to some driving force, the subsystems will
self-organize to a critical state at which a small perturbation can trigger
an avalanche-like chain reaction of any size (namely the instability). For
the GRB prompt emission pulses studied in this work, the instability of the
SOC behavior is closely related to the jet composition (thermally or magnetically
dominant) and the central engine models (hyper-accreting black hole or
rapidly spinning magnetar).
If the jet is thermally dominated
\cite{1993ApJ...405..278M,1993MNRAS.263..861P,2018ApJ...866...13H,2018ApJ...860...72M,2019ApJ...882...26M} and produced by the neutrino annihilation from
the accretion disk around the black hole, the instability is likely to be
the thermal instability of the accretion disk. If the jet is dominated by
magnetic fields \cite{2011ApJ...726...90Z,2014ApJ...793...36L,2014NatPh..10..351U}
and launched from a magnetar or a black hole (i.e., through the Blandford-Znajek mechanism \cite{1977MNRAS.179..433B}), the SOC behavior may be determined by the magnetic reconnection instabilities, such as the
kink-mode or tearing-mode instability.

It is interesting to notice that both the GRB prompt emission and X-ray flare
resemble a SOC system, but favor a different spatial dimension parameter respectively.
%{\bf Note that X-ray flares may be physically different from the GRB pulse since the flares may reflect the late active episodes of the GRB central engine. They are also %composed of overlapped pulses. In this scenario, the prompt burst phase and flare episodes should belong to different active stages of the central engine (e.g., %\citep{2014ApJ...789..145H}). The GRB pulses, on the other respect, should be the random collisions of shells in the GRB fireball. This may explain the different SOC feature %in the pulse and X-ray flares.}
This apparent contradiction could be also naturally explained within the scenario of a magnetic-dominated jet.
Some studies have suggested that X-ray flares and the gamma-ray prompt emission may share a common origin,
i.e., X-ray flares also come from relativistic jets. In such a scenario, the magnetic reconnection is supposed to drive the current sheets, from which the electrons are accelerated to emit observed photons.
However, the prompt burst phase and flare episodes should belong to different active stages of the central engine (e.g.,~\citep{2014ApJ...789..145H}).
During the early prompt phase, the turbulent magnetic reconnection will produce mini-emitters \cite{2014ApJ...782...92Z,2020PhPl...27a2305L}
by the magnetic instabilities (tearing/plasmoids). The runaway growth of these mini-emitters is in a 3-dimensional (or isotropic) form due to the existence of turbulence,
resulting in a SOC system with a dimension of $S = 3$. For late X-ray flares, the magnetic field
reconnection topology for the dissipation region should be 1-dimensional (i.e., $S = 1$) \cite{2013NatPh...9..465W}. The emission site of gamma-ray emission differs from that of X-ray flares, which makes the dimension of their magnetic reconnection different, i.e., the bursts in the prompt emission might correspond to 3-dimensional magnetic reconnection, while the X-ray flares might be 1-dimensional magnetic reconnection.
Therefore, our statistical results provide an indirect clue to identify the jet composition and the radiation mechanism of GRBs.

In this work, the GRBs are selected from the same instrument, and the backgrounds are well deducted when deriving the lightcurve of every burst, which makes the observational bias of our sample to be minimal. On the other hand, it seems to be a critical issue when we are gathering the pulses from different GRBs
to perform analyses while these GRBs may have different physical progenitors. In principle, it is better to analyze the pulses within a single burst. However, we found that
there are no enough pulses to perform the statistics within a single burst.
Fortunately, the selection criterion of our sample, i.e., having $\ge$ 3 pulses in each burst,
excludes any short GRB from our sample. All the bursts in our sample are  long GRBs indeed, and they are theoretically believed to come from the collapse of the massive stars.
Thus the bias of our results due to the internal difference between each burst is not so significant given the same origin.

To summarize, we draw a tentative conclusion that our statistical results are explained in the theoretical prediction of a self-organized criticality system with the classical diffusion, the spatial dimension $S = 3$, which are generally consistent with statistical results of various black hole systems including GRBs, TDE Swift J1644+57, Sgr A$^{*}$, M87. They can be explained by a three-dimensional SOC model \citep{2015ApJS..216....8W}, despite the dimension of our result is different from that of previous work \citep{2013NatPh...9..465W,2015ApJS..216....8W,2016ApJS..224...20Y,2017ApJ...844...79Y}, which implies that the relativistic jets may be magnetically dominated, consistent with our previous work \citep{2014ApJ...793...36L}.

Although only the BATSE sample of GRBs with pseudo redshifts are adopted for SOC analyses, we anticipate that the Swift sample of GRBs with detected redshifts and corresponding analysis may be also interesting. In the future, it is worthy to explore the SOC behavior in the prompt emission and X-ray flares simultaneously observed with a large sample of GRBs detected by Swift satellite, which may help to probe into the activities of the central engine from the prompt phase to afterglow phase.

\acknowledgements{
This work is partially supported by the National Natural Science Foundation of China (grant Nos.
11673068, 11725314, U1831122, 11703064,  11903019, and U1938116),
the Shanghai Sailing Program (No. 17YF1422600), the Youth Innovation Promotion Association (2017366),
the Key Research Program of Frontier Sciences (grant Nos. QYZDB-SSW-SYS005 and ZDBS-LY-7014),
and the Strategic Priority Research Program ``Multi-waveband gravitational wave Universe'' (grant No. XDB23000000) of the Chinese Academy of Sciences.		

}

%\bibliographystyle{fop}
%\bibliography{library,books}

%%%%%%%%%%%%%%%%%%%%%%%%%%%%%%%%%%%%参考文献的排法%%%%%%%%%%%%%%%%%%%%%%%%%%%%%%%%%%%%
%%%%%%%%%%%%%%%%%%%%%%%%%%%%%%%%%%%%注：\href{}{}%%%第一个括号表示需要链接的网址,第二个花括号表示锭接网址的内��?%%%%%%%%%%%%%%%
\raggedend
%\input{ref.bbl}
%\begin{multicols}{2}

\begin{small}

\end{small}

%\end{multicols}

\end{document}